\documentclass[prb,aps,amsmath,showpacs,preprint]{revtex4-1}

\usepackage{graphicx}
\usepackage[version=3]{mhchem}
\usepackage{dcolumn}
\usepackage{multirow}
\usepackage{color}
\usepackage{bm}
\usepackage{subfigure}
\begin{document}

\title{Calculated high-pressure structural properties, lattice dynamics and quasi particle band structures of perovskite fluorides
KZnF$_3$, CsCaF$_3$ and BaLiF$_3$}

\author{G. Vaitheeswaran$^{1*}$, V. Kanchana$^{2}$, Xinxin Zhang$^{3}$,
Yanming Ma$^{3}$, A. Svane$^{4}$, N. E. Christensen$^{4}$}
\affiliation{$^{1}$Advanced Centre of Research in High energy Materials (ACRHEM),
University of Hyderabad, Prof. C. R. Rao Road, Gachibowli, Hyderabad 500 046, Telengana, India \\
$^{2}$Department of Physics, Indian Institute of Technology Hyderabad, Kandi, Medak 502 285, Telengana, India \\
$^{3}$State Key Lab of Superhard Materials, Jilin University,
Chaungchun 130012, People's Republic of China \\
$^{4}$Department of Physics and Astronomy, Aarhus University,
DK-8000, Aarhus C,
Denmark}

\date{\today}

\begin{abstract}
A detailed study of the high-pressure structural properties, lattice dynamics and band structures of perovskite
structured fluorides KZnF$_3$, CsCaF$_3$ and BaLiF$_3$ has been carried out by means of density functional theory.
The calculated structural properties including elastic constants and equation of state agree well with available experimental information.
The phonon dispersion curves are in good agreement with available experimental inelastic neutron scattering data.
The electronic structures of these fluorides have been calculated using the quasi particle self-consistent $\it GW$  approximation. The $\it GW$  calculations reveal that all the fluorides studied are wide band gap insulators,
and the band gaps are significantly larger than those obtained by the standard local density approximation,
thus emphasizing the importance of quasi particle corrections in perovskite fluorides.
\end{abstract}


\maketitle

\section{Introduction}
 The large class of perovskite compounds, generally expressed as ABX$_3$, where A and B are cations and X is an anion,
have received great attention from experimentalists and theoreticians due to their interesting properties
 from the fundamental physics and chemistry point of view. They possess a wide range of applications such
as lenses without birefringence, and exhibit magnetism, pizeoelectrics, ferroelectrics etc.\cite{perovskites}
The physical properties of perovskites may also have implications towards understanding the Earth's lower mantle.\cite{keeffe,street}
  Fluoride perovskites is one particular subclass, which has found important technological applications in the field of optics.
	The main advantage is that these materials can be used as light emitting materials in the
deep ultraviolet regime.\cite{yamanoi} In the present work the three technologically important fluoroperovskite
KZnF$_3$, CsCaF$_3$ and BaLiF$_3$ are considered. They all crystallize in the cubic perovskite structure. KZnF$_3$ is
a promising candidate for application as a radiation detector.\cite{mortier1994}
Several experimental studies focused on synthesis, lattice dynamics, high-pressure structural stability,
elastic constants,
optical absorbtion and photoluminescence.\cite{lee, young, rousseau1981, lehner, burriel, aguado, tyagi}
In addition, several theoretical studies of KZnF$_3$ investigated its electronic structure and optical properties,
lattice dynamics and thermodynamic properties, crystal fields and influence of 3$d$ transition metals
dopants.\cite{khenata, fernandez, salaun, meziani}
Similar to KZnF$_3$, the CsCaF$_3$ perovskite also was investigated in several experimental and theoretical studies
addressing the electronic structure and optical properties, low energy phonon dispersion curves,
green luminescence upon Eu$^{2+}$ doping, thermodynamics at low temperatures, and the equation of
state.\cite{rousseau1981, sommerdijk, felix, boyer, murtaza, ephraim, meziani2012, brik}

  BaLiF$_3$ has been of interest for application within lithography due to the short wavelength absorption edge at 123 nm
(10.1 eV).\cite{sato2002}
In addition, BaLiF$_3$ may also be used as an effective dopant to alter the dielectric properties
of BaTiO$_3$.\cite{benziada1984}
The mechanism of ionic conductivity has also been investigated in BaLiF$_3$ with the purpose of enhancing the
 ionic conductivity by means of intrinsic defects.\cite{zahn}

The infrared dielectric dispersion of BaLiF$_3$ measured at several temperatures confirmed that this compound is one of the
most stable fluoroperovskites.\cite{boumriche1989}
This was further confirmed by inelastic neutron scattering measurements of the phonon spectrum.\cite{boumriche1994}
The electron-phonon coupling in Ni$^{2+}$ doped BaLiF$_3$ has been measured and related to the
change in bonding as manifested in the computed vibronic spectra.\cite{mortier}
The luminescence of BaLiF$_3$:Eu$^{2+}$ has been investigated under pressure.\cite{mahlik}
High pressure studies on fluoride compounds has resulted in various interesting physical properties. 
A recent high pressure experimental and theoretical studies on fluoride compound
CaF$_2$ reported superionic behavior even at ambient conditions which decreases with compression.\cite{daniel}
Recently, the structural stability of cubic BaLiF$_3$ under 
pressure has been confirmed experimentally up to 
50 GPa.\cite{mishra}
Similar results have been reported recently for the perovskite fluorides KMgF$_3$ and CsCdF$_3$ using high-pressure
experimental techniques up to 40 and 60 GPa, respectively.\cite{vaitheekmgf3, vaitheecscdf3} \emph{i.e.,}
the cubic phases of KMgF$_3$ and CsCdF$_3$ are found to be stable in the pressure ranges studied.

  The present paper reports the results of a
systematic and comparative study of the structural properites, lattice dynamics and electronic structures, of the three
perovskite fluorides KZnF$_3$, BaLiF$_3$ and CsCsF$_3$.
 The remainder of the paper is organised as follows.
Section II describes the computational details. The calculated ground state properties, elastic constants,
and phonon dispersion curves are discussed in section III, and section IV presents the
quasiparticle band structures of the fluoroperovskite compounds. Conclusions of the work are presented in section V.

\section{Computational details}
  Three complementary density functional tools have been employed to study the ground state and elastic properties,
lattice dynamics, and quasiparticle band structures of KZnF$_3$, CsCaF$_3$ and BaLiF$_3$.
The structural and elastic properties were calculated using the all-electron linear muffin-tin orbital method\cite{oka} in the full-potential implementation of Ref. \onlinecite{savrasov}. In this method the crystal volume is divided into two regions: nonoverlapping muffin-tin spheres surrounding each atom and the interstitial region between the spheres. We used a double $\kappa$ spdf LMTO basis
set to describe the valence bands. The calculations include $3s$, $3p$, $4s$, $4p$ and $3d$ partial waves for potassium,
$4s$, $4p$ and $3d$ partial waves for zinc, $5s$, $5p$, $6s$ and $5d$ partial waves for cesium,
$3s$, $3p$, $4s$, $4p$, $4d$ partial waves for calcium, $5s$, $5p$, $6s$, $5d$ and $4f$ partial waves for Ba,
and $2s$ and $2p$ partial waves for fluorine. The exchange correlation potential was calculated within the local density approximation (LDA)\cite{vosko}  as well as the generalized gradient approximation (GGA) scheme.\cite{perdew}
The charge density and electron potential inside the muffin-tin spheres were expanded in terms of spherical harmonics
up to $l_{max}$=6, while in the interstitial region, these quantities were expanded in plane waves, with
 28670 waves (energy up to 194.81 Ry)  being included in the calculation.
Total energies were calculated as a function of volume, for a $(16\times 16\times 16)$ k-mesh,
corresponding to 165 k-vectors in the irreducible wedge of the Brillouin zone (BZ),
and the results fitted to the Birch equation of state\cite{birch} to obtain the ground state properties. The elastic constants were obtained from the variation of the total energy under volume-conserving strains,
as outlined in Ref. \onlinecite{oxides}.

The phonon frequencies were calculated based on the density functional linear-response approach\cite{giannozzi} within
the GGA,\cite{perdew} combined with the plane-wave pseudopotential method as implemented in the PWSCF package.\cite{baroni}
Norm-conserving pseudopotentials were used to describe the valence electrons. The BZ integrations were carried out with a
	$(6\times 6\times 6)$
 Monkhorst-Pack (MP) grid.\cite{monkhorst} Kinetic energy cutoffs of 90 Ry, 90 Ry and 120 Ry were used for
CsCaF$_3$, BaLiF$_3$ and KZnF$_3$, respectively. A
	$(4\times 4\times 4)$
 $q$ mesh in the first BZ was used for the interpolation of the
force constants needed for the phonon dispersion curve calculations.
The theoretical equilibrium lattice constant of 4.096 \AA, 4.604 \AA, and 4.085 \AA\ for KZnF$_3$, CsCaF$_3$ and BaLiF$_3$, respectively, are obtained by total energy minimization and used in the phonon calculations.


The electronic structures of KZnF$_3$, CsCaF$_3$ and BaLiF$_3$ have been
calculated with the quasiparticle self-consistent $GW$ (QSGW) (where
$G$ denotes the Greens function and $W$ denotes the screened Coulomb
interaction) approximation,\cite{Kotani1,kotani} in the FP-LMTO implementation
of Ref. \onlinecite{methfessel}.
This calculation also included a
double-$\kappa$ LMTO basis set ($\ell_{max}=6$), however additionally including the F $3s$ and $3p$ partial waves
(treated as local orbitals\cite{methfessel}) for better description of the unoccupied band states. An  $(8\times  8\times  8)$ MP
k-mesh was used for the calculation of the screened interaction,
which is evaluated in the random-phase approximation.
Convergence tests showed only minor changes compared to a $(6\times  6\times  6)$ mesh. The most common type of $GW$
calculations use a selfconsistent LDA or GGA band structure as input for the evaluation of the $G$ and $W$ operators.
The special feature of the quasiparticle self-consistency in the QSGW method is that
the one-particle band structure, which is used as input
for the evaluation of the Greens' function, is iterated in a further selfconsistency loop so as to
come as close as possible to the output GW band structure.\cite{Kotani1,kotani} Though appealing from a theoretical point of view, this procedure has
a tendency to slightly overestimate semiconductor gaps,\cite{Chantis} a fact traceable to the neglect of vertex corrections. As a simple remedy of this,
Chantis et al.\cite{Chantis} suggested the hybrid QSGW (h-QSGW), by which the
LDA and QSGW self energies are mixed in the proportions 20\% - 80\%. See Ref. \onlinecite{Chantis-ss} for a detailed description. Here we also apply the h-QSGW method to the fluorite perovskites to monitor its effect for this class of wide-gap insulators


\section{Ground state, elastic properties and phonon dispersions}


The calculated structural properties such as lattice parameters, and bulk moduli  of KZnF$_3$, CsCaF$_3$ and BaLiF$_3$  are presented
 in Table I along with available experimentalal information and results of other calculations.
 In the case of KZnF$_3$ the calculated lattice parameter within LDA
 is 0.9$\%$ lower when compared with the experimental value and the corresponding bulk modulus is 23$\%$ overestimated.
In contrast, the calculated lattice parameter using GGA overestimates the experimental value by 2.4$\%$, and the bulk modulus
is 6.8$\%$  too low. This is the usual trend observed in the LDA and GGA schemes.
In the case of CsCaF$_3$ the LDA calculated lattice constant is lower by 2.8$\%$ when compared to experiments, and the corresponding bulk modulus is 44$\%$ higher. Similarly, the GGA lattice parameter is 0.4$\%$ lower and the corresponding bulk modulus is 8\% too
large. A similar situation is seen in the case of BaLiF$_3$. The calculated bulk modulus is in excellent agreement
with the two reported experimental values, which is, however, a bit fortuitous. Since the calculated equilibrium volume is overestimated
(by about 0.8\%) with GGA (and underestimated by about 2.6 \% with LDA), an error solely depending on the error in volume is introduced in the calculated bulk modulus. Therefore we recalculated the bulk modulus also at the experimental volume in a manner similar to our earlier work\cite{vaitheekmgf3} (see Table I). This diminishes the discrepancies between the LDA and GGA results, as expected. In addition, the LDA bulk modulus now becomes $\it smaller$ than the GGA for all the three compounds, and both functionals are seen to actually overestimate the bulk modulus.
For KZnF$_3$  two experimental determinations of the bulk modulus are at variance.
In particular the high value for the pressure derivative  of the bulk modulus ($B_0'=11$) reported by
Ref. \onlinecite{aguado} is at odds with the general range $B_0'\sim 4-5$ found for the fluoride perovskites
by both experiments and theory. The authors suggest that the limited pressure range covered in
their study may be the cause of their high fit value for $B_0'$.
In Figure 1 the GGA and LDA equations of state of KZnF$_3$, CsCaF$_3$ and BaLiF$_3$ in the pressure range from
0 to 50 GPa are compared with the available experimental data.\cite{aguado, mishra}
The agreement between theory and experiment is excellent for BaLiF$_3$ in Fig. 1(c),
while the agreement is less satisfactory for
KZnF$_3$ in Figure 1(a). A special feature in KZnF$_3$ is the role of the Zn 3d states,
which occur as a narrow resonance within the F valence bands reflecting their semi-localized character.
Their bonding properties are usually ill described in LDA/GGA and this may be the cause of the discrepancy here.\\

The elastic constants of  KZnF$_3$, CsCaF$_3$ and BaLiF$_3$  are presented in Table II
along with the experimental values and other
theoretical values. 
The LDA overestimates all the C$_{11}$, C$_{12}$ and C$_{44}$ by between 10$\%$ and 25$\%$,
while the elastic constants obtained within GGA are much closer to the experimental values.
As for the bulk modulus the elastic constants also depend sensitively on the volume, and therefore,
the same argument as for the bulk modulus can be applied
here. We have however, refrained from recalculating all the elastic constants with volume correction,
but wish to mention that the
excellent agreement between experiment and the GGA elastic constants should be interpreted with some care.
Another point of caution
is the fact that the calculated values pertain to 0 K, while experiments are performed at room temperature.
An increase in temperature generally tends to reduce the elastic constants because of thermal expansion.\\

The calculated phonon dispersion curves at the optimized theoretical lattice parameters are shown in Figure 2.
The available experimental inelastic neutron scattering frequencies\cite{rousseau1981,boumriche1994} are included and good agreement between
theory and experiment is found. The dynamical stability of the cubic structures is confirmed by the absence of
any imaginary modes in the calculated dispersions. There are 15 phonon branches in the full phonon
dispersion since the unit cell consists of five atoms which
give rise to three acoustic and twelve optical phonon branches.
The agreement with experiment is excellent for BaLiF$_3$, and for the few frequencies determined for CsCaF$_3$, while there is some discrepancy to be seen for KZnF$_3$ around the lowest frequencies at the M and R points. The calculations find a noticeable
gap between the lowest optical mode and the highest acoustic mode in the calculations for BaLiF$_3$ and CsCaF$_3$, which does not appear in KZnF$_3$. In addition, in CsCaF$_3$ a large second gap appear at high frequency. The experiments have only managed to map the lowest frequencies and thus cannot confirm or discard this prediction.  Evidently, it would be of great interest to obtain experimental phonon frequencies in the higher end of the spectrum for all three compounds, in particular to test the theoretical prediction that the phonon spectrum
for  CsCaF$_3$, BaLiF$_3$ and KZnF$_3$ contain two, one and zero gaps, respectively.

The phonon partial densities of states are also included to the right of Figures 2(a)-(c). The acoustic modes in BaLiF$_3$ and
CsCaF$_3$ are dominated by the motion of the heavy Ba and Cs, respectively, while in KZnF$_3$ the atomic masses are more similar, and the atomic vibrations are fully coupled to each other as shown in figure 2(a). Furthermore, the acoustic band width is smaller in
CsCaF$_3$ than in BaLiF$_3$,  which reflects the larger lattice constant of CsCaF$_3$, while the second gap (between 8 and 11 THz) in the phonon spectrum of CsCaF$_3$, which does not find a counterpart in BaLiF$_3$, reflects the different masses of Ca and Li.
Thus, there is a significant Li component to the highest frequencies in BaLiF$_3$, while no Ca weight is found in the highest   CsCaF$_3$ frequency range.

\section{Quasiparticle Band Structures}

The QSGW band structures of KZnF$_3$, CsCaF$_3$ and BaLiF$_3$  are presented in Figure 3. All three compounds exhibit large gaps,
of 10.0 eV, 11.8 eV and 11.8 eV, respectively, which are significantly larger than the gaps calculated within the LDA (Table III).
The experimental
gap has only been reported for the case of BaLiF$_3$ to be 8.4 eV,\cite{yamanoi} which is 3.4 eV lower than the calculated QSGW value. The h-QSGW
reduces the calculated gaps by about 1 eV (Table III), i. e. this can not account for the discrepancy.
There exists the possibility that the ideal perovskite structure assumed in the calculations is not the same as that of the crystals grown in the laboratory,
which may be hampered by intrinsic defects etc.

The conduction band minimimum  occurs in all cases at the $\Gamma$ point, while the valence band maximum (VBM) also falls at $\Gamma$
for  BaLiF$_3$, \emph{i.e.,} this is a direct gap insulator.  The three top valence states at $\Gamma$, M and R are rather close, with M (R) at -0.17 eV (-0.20 eV) relative to the VBM in BaLiF$_3$.
In contrast, for CsCaF$_3$ and KZnF$_3$ the gap is  indirect.
In CsCaF$_3$ the VBM occurs at the M point while the topmost valence state at $\Gamma$ is about 0.25 eV lower in energy. In KZnF$_3$
 the VBM occurs on the line connecting the M and R points,
and the topmost valence state at $\Gamma$ is 1.5 eV lower in energy.

The F band widths are 4.0 eV, 2.1 eV and 3.2 eV in KZnF$_3$, CsCaF$_3$ and BaLiF$_3$, respectively.
The low band width of CsCaF$_3$ reflects its larger lattice constant.
KZnF$_3$ and BaLiF$_3$ have similar lattice constants,
however in KZnF$_3$ the Zn 3d bands falls at the bottom of the F bands partially hybridizing
and thus effectively enhancing the F band width.
The semicore states are slightly shifted towards higher binding energies by the GW approximation compared to their LDA positions,
by about 2 eV for the K 3p states of  KZnF$_3$ (located at 12 eV below VBM),
by about 0.1 eV for the Cs 5p states in CsCaF$_3$ (located at 4.5 eV below VBM),
and by about 0.6 eV for the Ba 5p states of BaLiF$_3$ (located at 9 eV below VBM).
Thus the most significant effect of the GW approximation is the more or less rigid shift of the conduction bands with respect to the valence bands, which is of the order of 5-7 eV for the three fluorides considered.

\section{Conclusions}

The electronic structures and vibrational properties of the KZnF$_3$, CsCaF$_3$ and BaLiF$_3$ perovskite fluorides have been investigated with density functional methods. The calculated phonon frequencies are in good agreement with experimental values, which however are limited to the lowest part of the spectrum, and a full experimental mapping of the phonon dispersion curves is desirable.
The calculated pV-curve of BaLiF$_3$ is in excellent agreement with experiment, while the agreement between theory and experiment is less satisfactory for KZnF$_3$, and no experimental data exist for CsCaF$_3$.

The QSGW quasiparticle band structures reveal significant insulating gaps in the fluorides studied, which are about 6 eV larger
than predicted by the LDA/GGA. Most of this effect may be seen as a rigid shift of the conduction bands with respect to the valence bands.
The F valence bands are only slightly wider in QSGW than in LDA, while some shifts towards higher binding energies are
found for the secondary (semi-core) valence states. All of these effects are usual trends of the GW method, which are also observed in comparison of Hartree-Fock and LDA band structures, hence they may be ascribed to the non-local character of the electron-electron
interaction.

\section{Acknowledgements}
G.V acknowledges, CMSD-University of Hyderabad for providing the computational facility. \\ $^*$\emph{Author for Correspondence, E-mail: vaithee@uohyd.ac.in}

\newpage

\begin{table}[tb]
 \caption{
Calculated Lattice constants (in \AA), bulk moduli $B_0$ (in GPa)
 and its pressure derivatives $B_0'$,
 of KZnF$_3$, CsCaF$_3$ and BaiF$_3$. The bulk moduli have been calculated both at the experimental and theoretical volumes
($B_0(V^{\rm{exp}}_{0})$ and $B_0(V^{\rm{th}}_{0})$, respectively).
The experimental bulk moduli reported for KZnF$_3$, CsCaF$_3$ and BaLiF$_3$ are obtained from high pressure structural
measurements (Refs. \onlinecite{aguado} and \onlinecite{mishra}), or derived
from the elastic constants: $B=(C_{11}+2C_{12})/3$ (Refs. \onlinecite{gesland1972}, \onlinecite{ridou1986}
and \onlinecite{boumriche1994}).}
\begin{ruledtabular}
\begin{tabular}{cccccc}
Compounds& Lattice constant &   $B_0(V^{\rm{th}}_{0})$  & $B_0(V^{\rm{exp}}_{0})$     & $B_0'                 $  \\
\hline
KZnF$_3$	\\

       LDA$^a$ & 4.017    & 95.2    &  83.5      &  4.4  \\
       GGA$^a$ & 4.154   & 72.3   &   99.8      &  4.3 \\
       Expt.   & 4.054$^b$, 4.060$^c$ &       & 77.6$^b$, 54.8$^c$   &  11$^c$ \\
      Other theory   & 4.021$^d$, 4.072$^e$, 4.07$^f$,  & 89.2$^d$ &  & 4.51$^d$\\
			               & 3.796$^g$, 3.897$^h$, 4.1499$^i$  &  66.4$^i$     &   & 4.22$^i$  \\

\\

CsCaF$_3$	\\

       LDA$^a$ & 4.401     & 73.5   &   47.8     &  4.6  \\
       GGA$^a$ & 4.545     & 55   &    58.2   &  4.5 \\
			 Expt.   & 4.526$^j$, &       & 50.9$^j$   &    \\
       Other theory   & 4.569$^k$, 4.579$^l$, 4.5885$^m$ & 50.2$^k$, 50.9$^l$, 45.8$^m$      &   &  4.27$^k$, 4.6$^m, $3.8$^l$  \\
\\

BaLiF$_3$	\\

       LDA$^a$ & 3.892     & 98.2    &  65.8      &  4.6  \\
       GGA$^a$ & 4.028   & 75.6   &   80.7     &  4.5 \\
       Expt.   & 3.995$^n$, &       & 79(3)$^n$, 75.9$\pm$1.3$^o$   &  5.35$\pm$0.15$^o$  \\
 			 Other theory & 4.050$^p$    & 64.5$^p$    &         &  4.6$^p$  \\

\end{tabular}
\end{ruledtabular}
$^a$Present work;
$^b$Ref. \onlinecite{gesland1972};
$^c$Ref. \onlinecite{aguado};
$^d$Ref. \onlinecite{khenata};
$^e$Ref. \onlinecite{jiang2006};
$^f$Ref. \onlinecite{moreira2007};
$^g$Ref. \onlinecite{verma2009};
$^h$Ref. \onlinecite{luana1997};
$^i$Ref. \onlinecite{brik2012};\\
$^j$Ref. \onlinecite{ridou1986};
$^k$Ref. \onlinecite{brik};
$^l$Ref. \onlinecite{meziani2012};
$^m$Ref. \onlinecite{ephraim};
$^n$Ref. \onlinecite{boumriche1994};
$^o$Ref. \onlinecite{mishra};
$^p$Ref. \onlinecite{korba2009}.

\end{table}

\newpage

\begin{table}[tb]
\caption{
Calculated elastic constants and  shear modulus (G), all expressed
in GPa, for KZnF$_3$, CsCaF$_3$ and BaLiF$_3$
at the theoretical equilibrium volume.}
\begin{ruledtabular}
\begin{tabular}{cccccc}
 Compounds             &$C_{11}$  &$C_{12}$  &$C_{44}$ &  $G$ &  \\ \hline
KZnF$_3$ \\
LDA           & 175.7     & 55.0     &  47.2             &  52.5   & Present   \\
GGA           & 134.8     & 41.1     &  44.1             &  45.2   & Present    \\
GGA           &  99.2     & 38.0     &  29.1             &  29.7   &  Ref.  \onlinecite{brik2012}   \\
GGA           & 111.8     & 49.8     &  31.4             &  31.2   &  Ref.  \onlinecite{meziani}   \\
Expt.         & 134.5$\pm$1  & 52.7$\pm$0.5  &  38.1$\pm$0.2  & 39.2 &  Ref.  \onlinecite{gesland1972} \\
Expt.         & 146          & 54            &  39            &      &  Ref.  \onlinecite{burriel} \\
\\

CsCaF$_3$ \\
LDA           & 166.2    & 27.3      &  29.2             &  45.3    & Present   \\
GGA           & 121.0    & 22.1      &  29.4             &  37.4    & Present    \\
GGA           &  98.5    & 24.8      &  27.6             &  31.0    & Ref.  \onlinecite{brik}    \\
GGA           &  99.9    & 23.6      &  25.0             &  29.6    & Ref.  \onlinecite{meziani2012}    \\
Expt.         & 102$\pm$1  & 25.3$\pm$0.5  &  25.5$\pm$0.5  & 30.6 &  Ref.  \onlinecite{ridou1986} \\

\\

BaLiF$_3$ \\
LDA           & 232.3     & 31.2      &  42.2              &  65.5    & Present   \\
LDA           & 163.8     & 50.8      &  50.0              &          & Ref.  \onlinecite{korba2009}   \\
GGA           & 149.2     & 33.1      &  58.0              &  58.0    & Present    \\
Expt.         & 130$\pm$1  & 46.5$\pm$0.5  &  48.7$\pm$0.5  & 45.9 &  Ref.  \onlinecite{boumriche1994} \\

\end{tabular}
\end{ruledtabular}
\end{table}

\newpage

\begin{table}[h]
\caption{
Comparison of the fundamental band gap (in eV), for KZnF$_3$, CsCaF$_3$ and BaLiF$_3$, as calculated with GGA
 and QSGW.
}
\begin{ruledtabular}
\begin{tabular}{ccccc}
   compound            &    LDA       & QSGW    &  h-QSGW      &  Expt.              \\
KZnF$_3$               &    2.2       &   10.0  & 8.7               &                     \\
CsCaF$_3$              &    6.1       &   11.8  & 10.7               &                     \\
BaLiF$_3$              &    6.3       &   11.8  & 10.7     &   8.41$^a$          \\

\end{tabular}
\end{ruledtabular}
$^a$Ref. \onlinecite{yamanoi}; 
\end{table}

\newpage

\begin{figure}

\subfigure[]{\includegraphics[width=70mm,height=70mm]{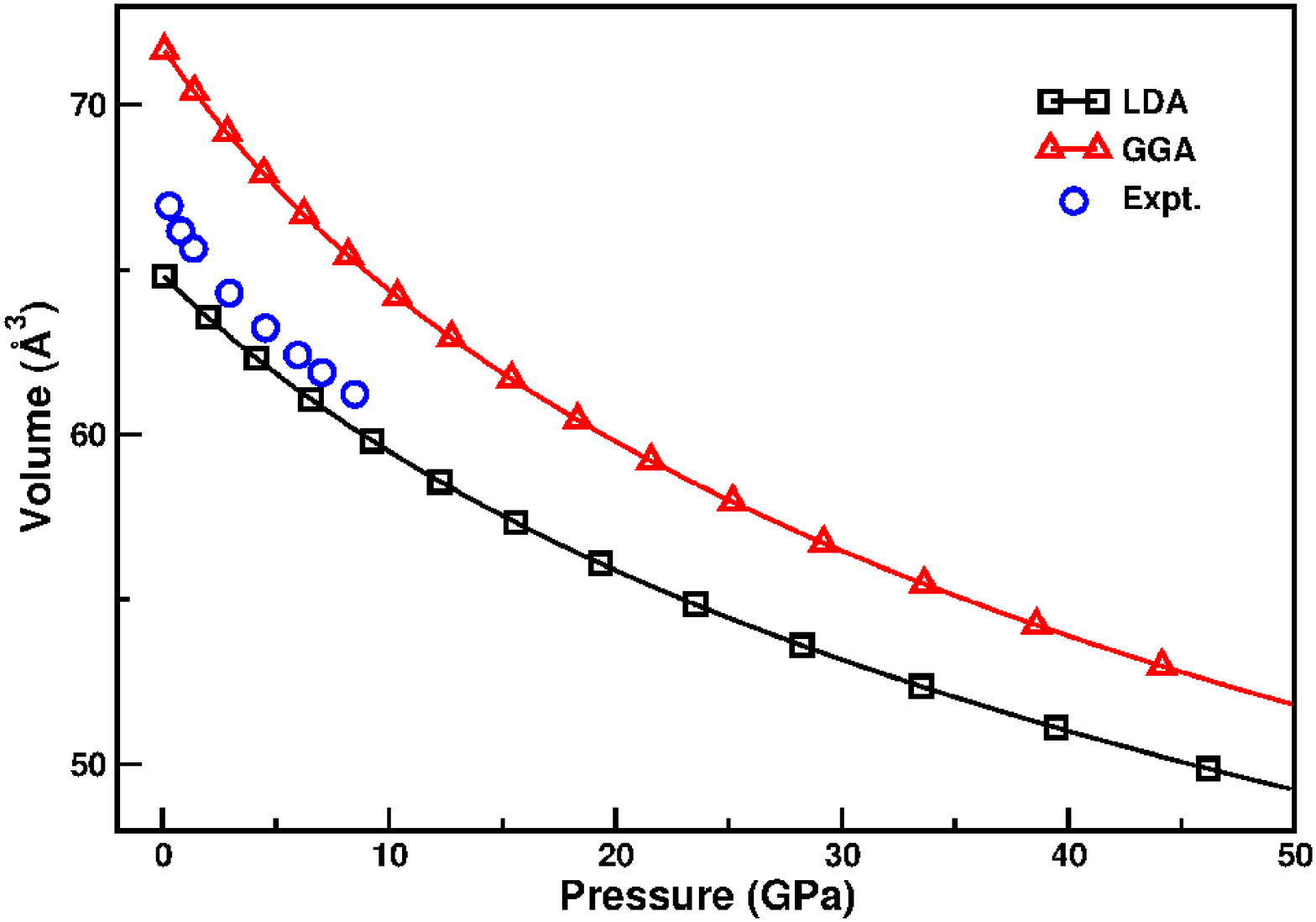}} \hspace{1cm}
\subfigure[]{\includegraphics[width=70mm,height=70mm]{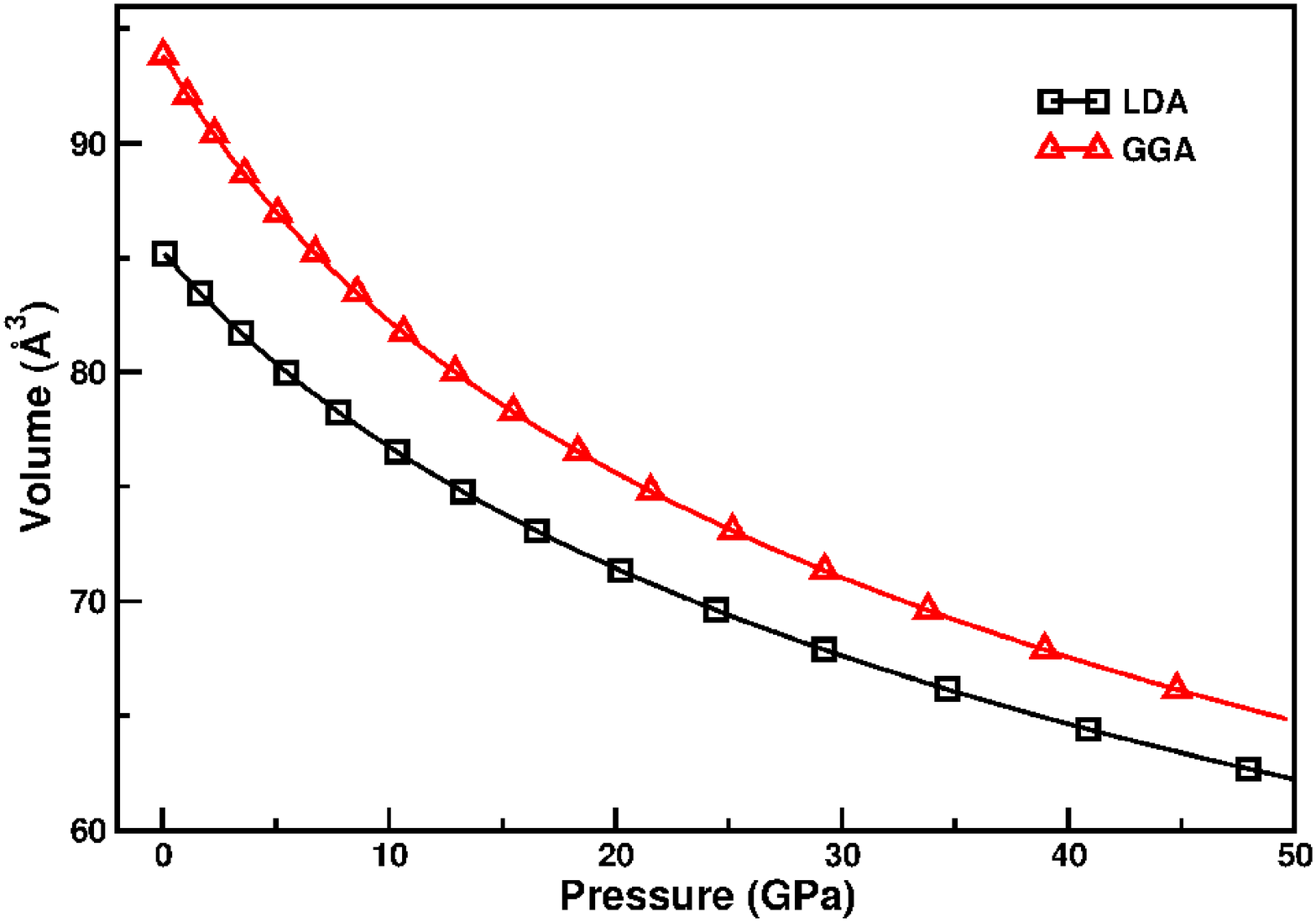}} \\[2cm]
\subfigure[]{\includegraphics[width=70mm,height=70mm]{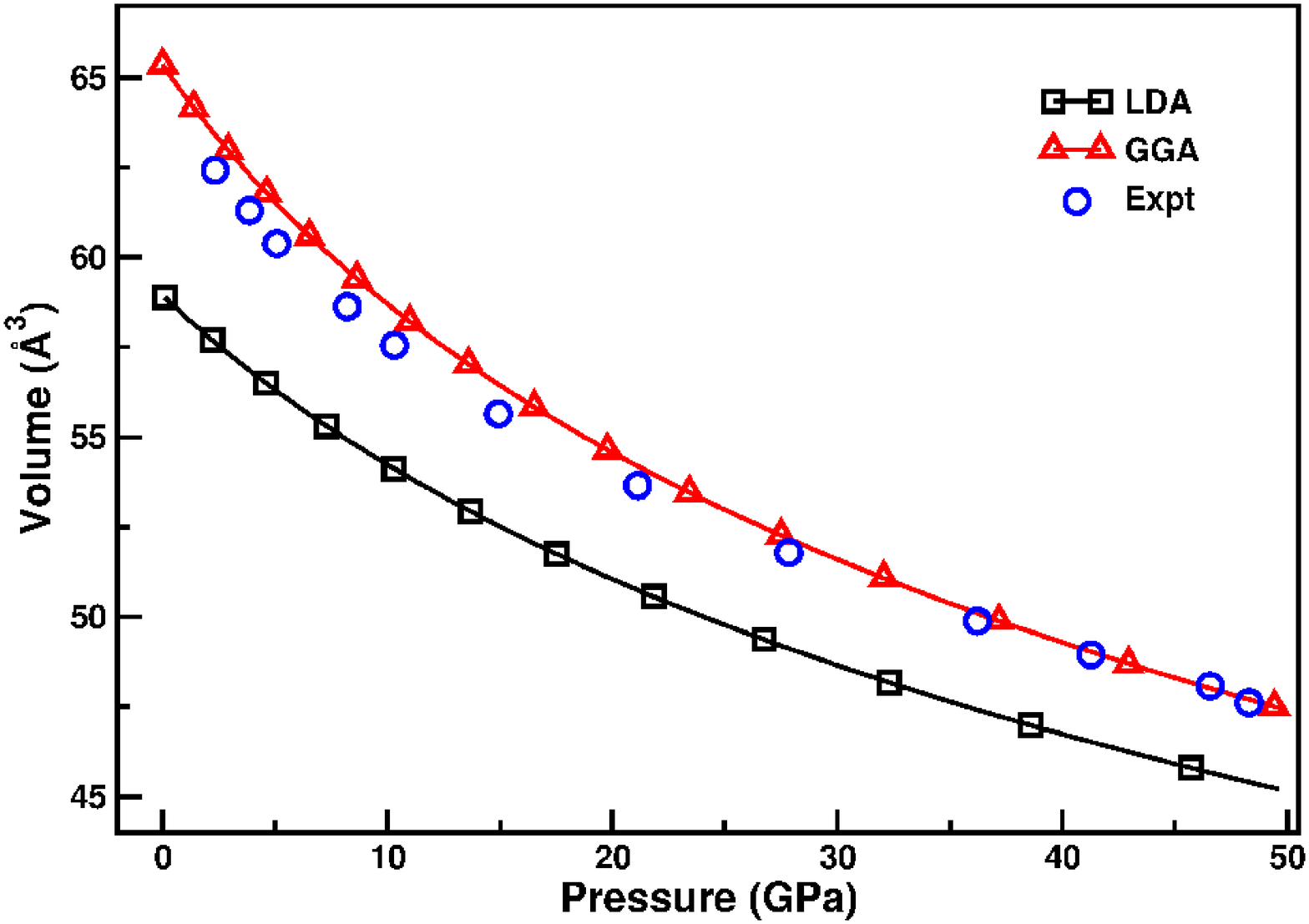}} \\

\caption{Calculated equation of state (LDA and GGA) versus experimental $pV$ relations for (a) KZnF$_3$,  (b) CsCaF$_3$ and
(c) BaLiF$_3$. Experimental data are from Refs. \onlinecite{aguado} (KZnF$_3$) and \onlinecite{mishra} (BaLiF$_3$).
}
\end{figure}

\newpage




\begin{figure}

\subfigure[]{\includegraphics[width=90mm,height=60mm, angle=360]{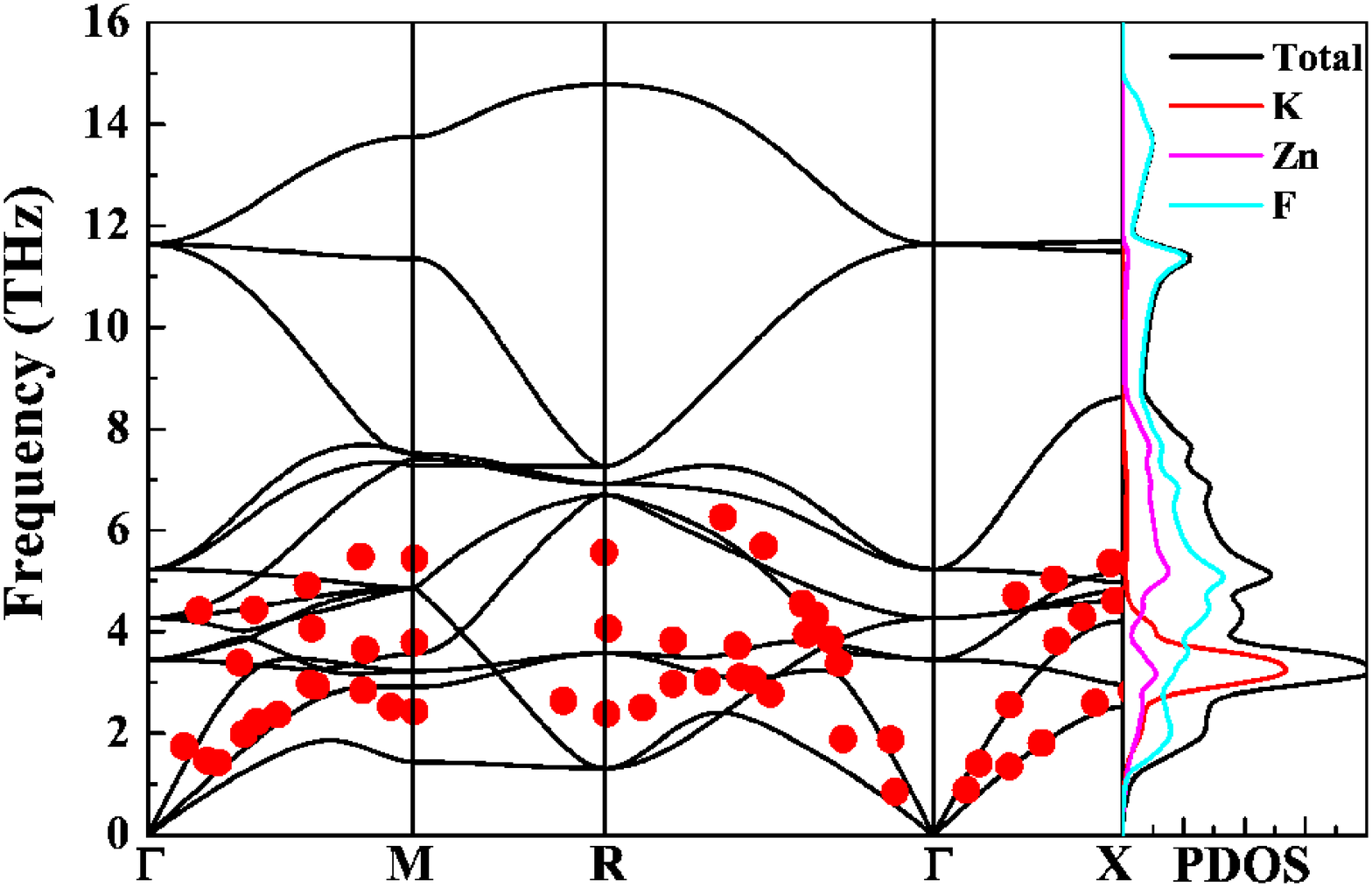}}
\subfigure[]{\includegraphics[width=90mm,height=60mm, angle=360]{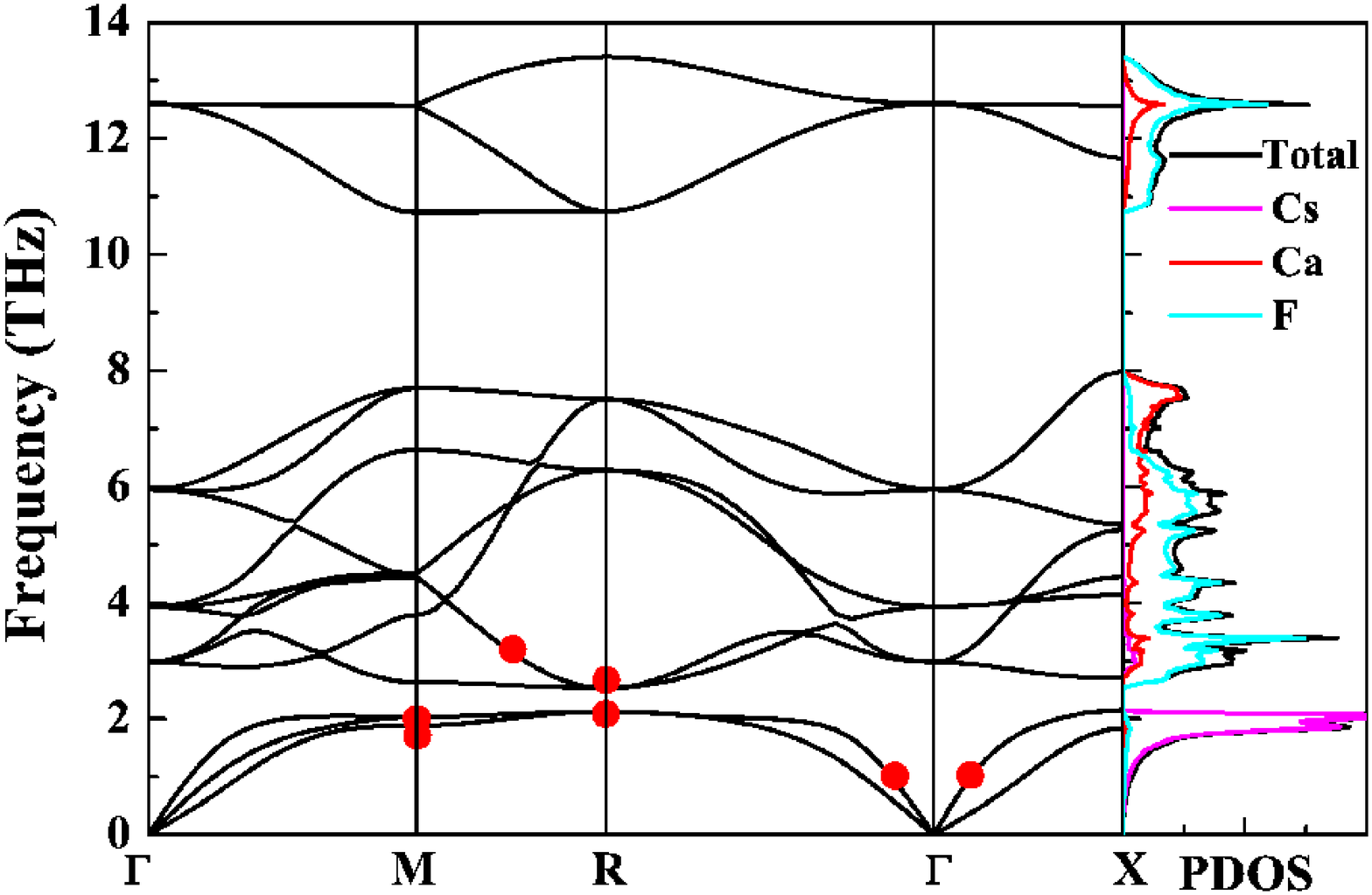}}
\subfigure[]{\includegraphics[width=90mm,height=60mm, angle=360]{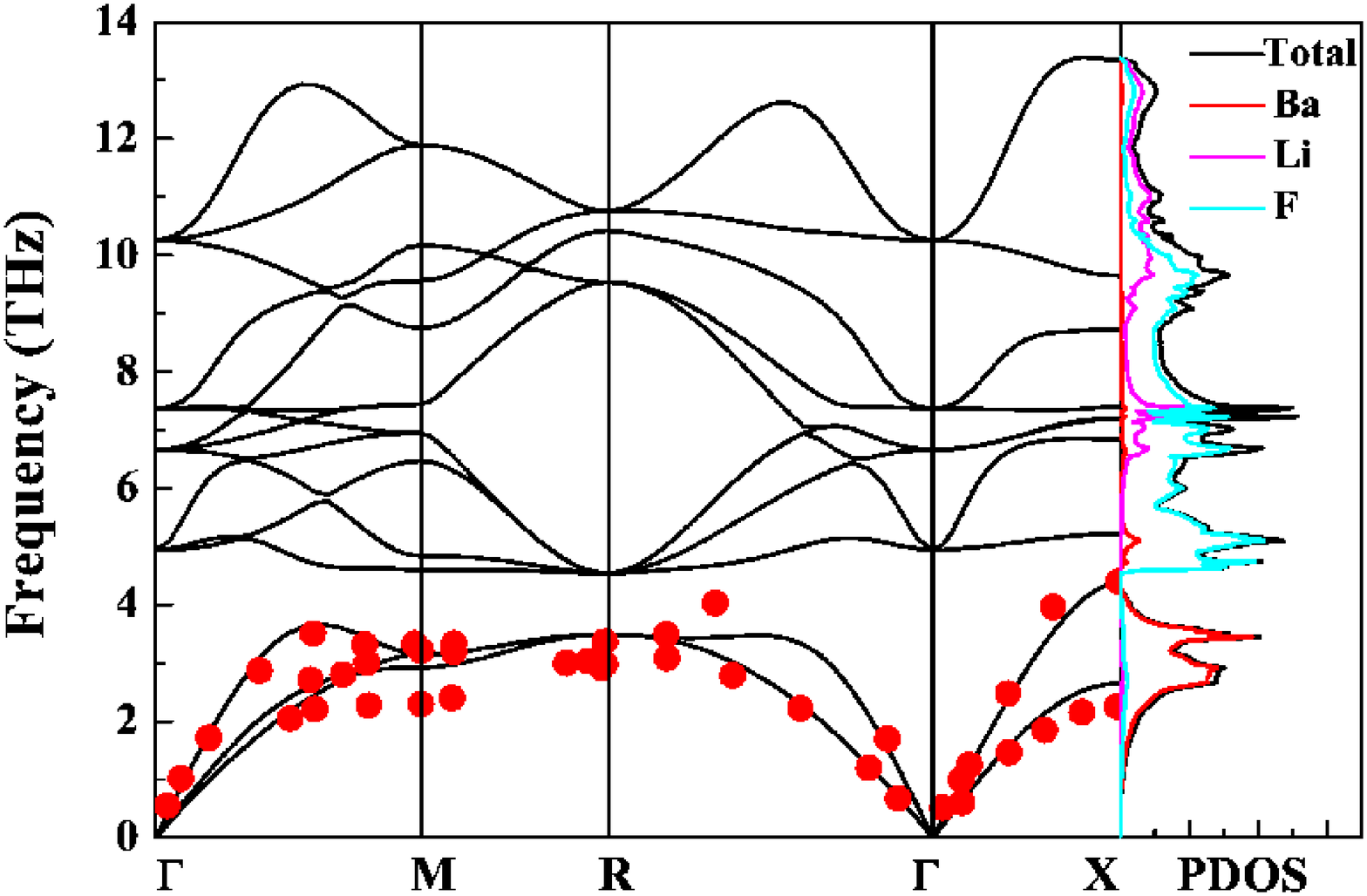}}\\

\caption{Calculated phonon dispersion curves of  (a) KZnF$_3$ and (b) CsCaF$_3$ and (c) BaLiF$_3$ at  the theoretical lattice parameter. The experimental data are shown by dots and taken from:  KZnF$_3$: (Ref. \onlinecite{rousseau1981});
CsCaF$_3$: (Ref. \onlinecite{rousseau1981});  BaLiF$_3$: (Ref. \onlinecite{boumriche1994}).
}
\end{figure}


\begin{figure}

\subfigure[]{\includegraphics[width=70mm,height=50mm, angle=360]{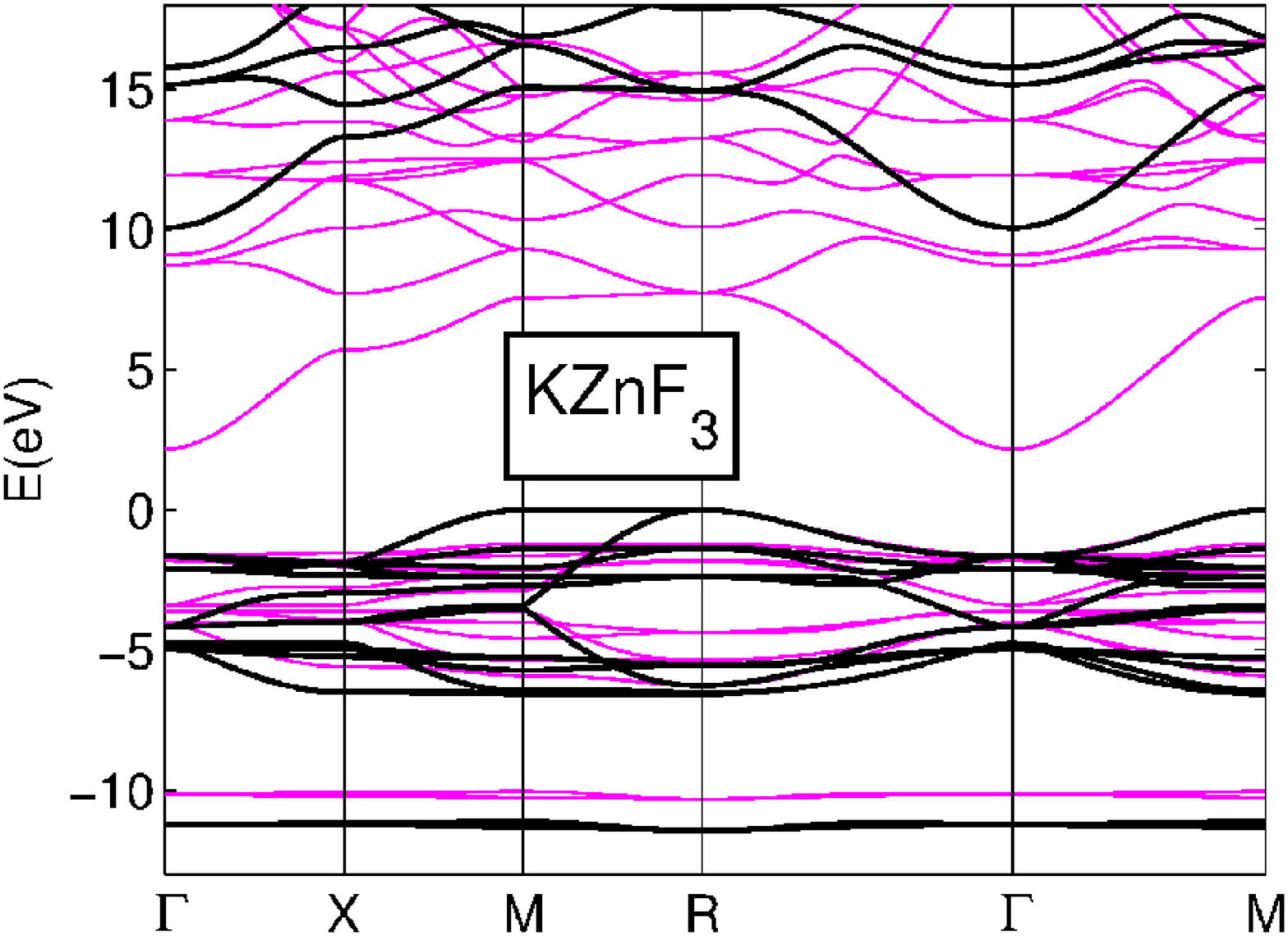}}\\
\subfigure[]{\includegraphics[width=70mm,height=50mm, angle=360]{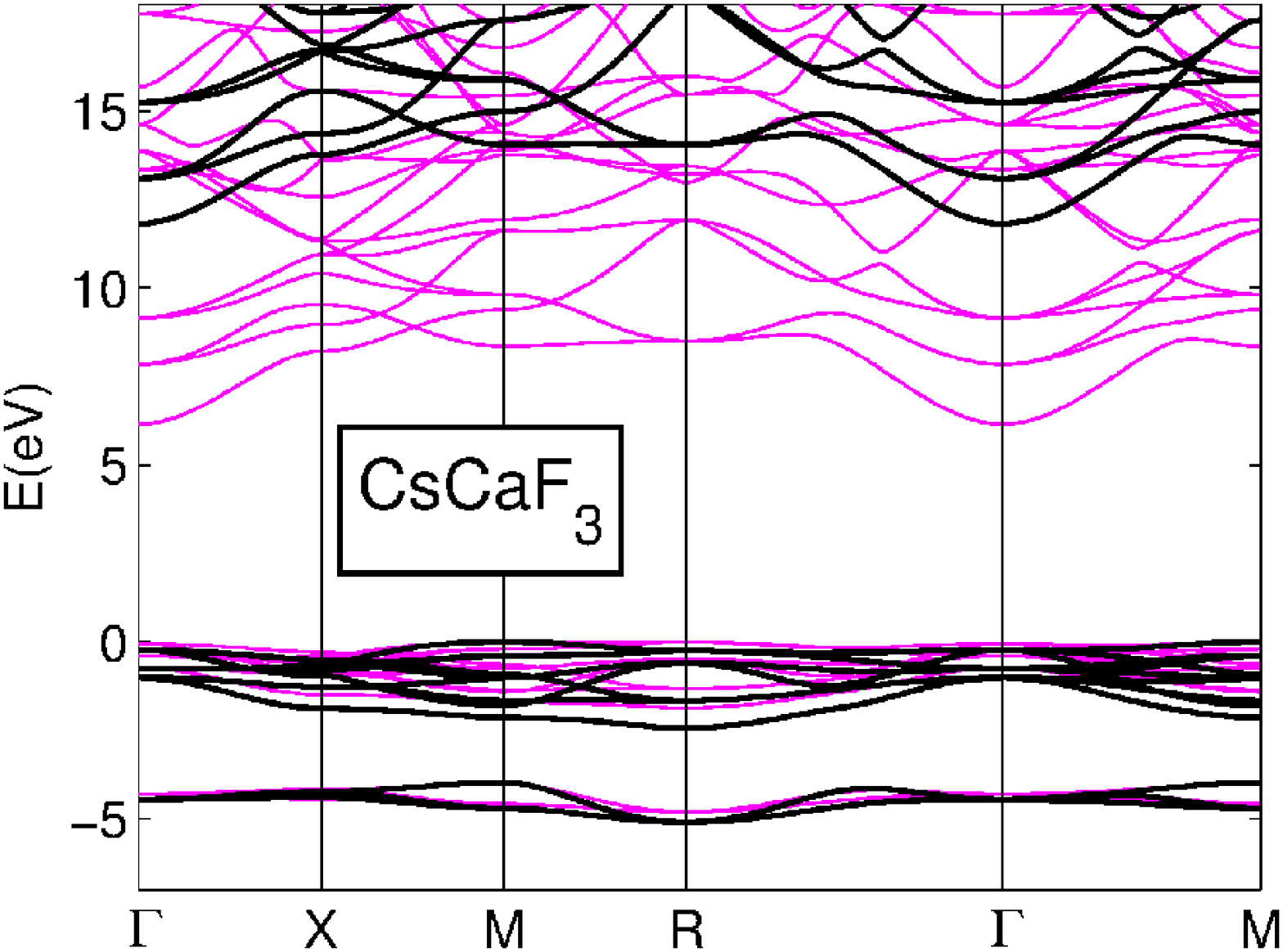}}\\
\subfigure[]{\includegraphics[width=70mm,height=50mm, angle=360]{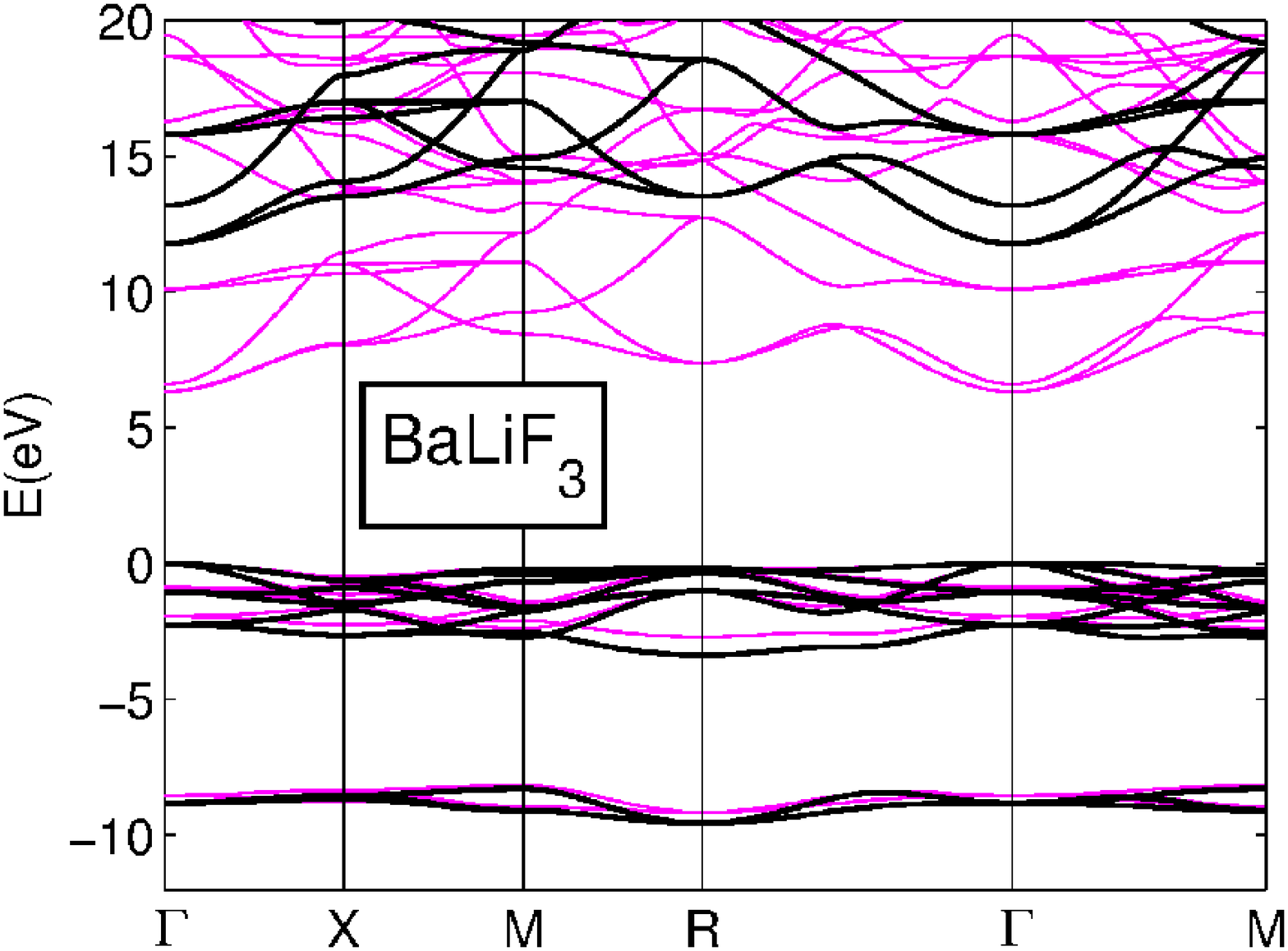}}\\

\caption{(Color Online) Calculated LDA  (red (gray thin) lines) and QSGW (black thick lines)  band structures of (a) KZnF$_3$, (b) CsCaF$_3$, and
(c) BaLiF$_3$ at the experimental lattice constants. The zero of energy is situated at the valence band maximum. The BZ
special points are $\Gamma$: (0,0,0); X: (1,0,0); M: (1,1,0) and R: (1,1,1) in units of $2\pi/a$.}

\end{figure}


\newpage

\begin{thebibliography}{}

\bibitem{perovskites}
        \emph{Perovskites: A structure of great interest to Geophysics and Material Science},
        edited by A. Navrotsky and D. J. Weidner, Geophys. Monogr. Ser., vol. {\bf 45}, 146.,
         American Geophysical Union, Washington DC (1989).
				
\bibitem{keeffe}
        M. \'O. Keeffe and J.-O. Bovin, Science, {\bf 206}, 599 (1979).

\bibitem{street}
        J. N. Street, I. G. Wood, K. S. Knight and G. D. Price,
        J. Phys.: Condens. Matter {\bf 9}, L647 (1997).

\bibitem{yamanoi}
        K. Yamanoi, R. Nishi, K. Takeda, Y. Shinzato, M. Tsuboi, M. V. Luong,
        T. Nakazato, T. Shimizu, N. Sarukura, M. Cadatal-Raduban, M. H. Pham, H. D. Nguyen,
				S. Kurosawa, Y. Yokota, A. Yoshikawa, T. Togashi, M. Nagasona  and T. Ishikawa,
				Optical Materials {\bf 36}, 769 (2014).
				
\bibitem{mortier1994}
        M. Mortier, J. Y. Gesland and M. Rousseau,
        Solid State Commun. {\bf 89}, 369 (1994).

\bibitem{lee}
        J. Lee, H. Shin, J. Lee, H Chung, Q. Zhang and F. Saito,
        Materials Transactions {\bf 44}, 1457 (2003).
				
\bibitem{young}
        E. F. Young and C. H. Perry, J. Appl. Phys. {\bf 38}, 4624 (1967).

\bibitem{rousseau1981}
        M. Rousseau, J. Y. Gesland, B. Hennion, G. Heger and B. Renker,
        Solid State Commun. {\bf 38}, 45 (1981).

\bibitem{lehner}
        N. Lehner, H. Rauh, K. Strobel, R. Geick, G. Heger, J. Bouillot, B. Renker, M. Rousseau and W. G. Stirling,
        J. Phys.: Condens. Matt. {\bf 15}, 6545 (1982).
				
\bibitem{burriel}
        R. Burriel, J. Bartolom\'e, D. Gonz\'alez, R. Navarro, C. Ridou, M. Rousseau and A. Bulou,
        J. Phys.: Condens. Matt. {\bf 20}. 2819 (1987).

\bibitem{aguado}
        F. Aguado, F. Rodriguez, S. Hirai, J. N. Walsh, A. Lennie and S. A. T. Redfern,
        High Press. Res. {\bf 28}, 539 (2008).
				
\bibitem{tyagi}
        N. Tyagi, P. Senthilkumar, R. Nagarajan,
        Chem. Phys. Lett. {\bf 494}, 284 (2010).
				
\bibitem{khenata}
        T. Seddik, R. Khenata, O. Merabiha, A. Bouhemadou, S. Bin-Omran and D. Rached,
        Appl. Phys. A {\bf 106}, 645 (2012).
				
\bibitem{fernandez}
        P. Garc\'ia-Fern\'andez, A. Trueba, B. Garci\'a-Cueto, J. A. Aramburu, M. T. Barriuso and M. Moreno,
        Phys. Rev. B {\bf 83}, 125123 (2011).
				
\bibitem{salaun}
        S. Sala\"un and M. Rousseau, Phys. Rev. B {\bf 51}, 15867 (1995).

\bibitem{meziani}
        A. Meziani, D. Heciri and H. Belkhir, Physica B {\bf 406}, 3646 (2011).

\bibitem{sommerdijk}
        J. L. Sommerdijk and A. Bril,  J. Luminescence {\bf 10}, 145 (1975).

\bibitem{felix}
        F. Koussinsa and M. Diot, Thermochimica Acta, {\bf 216}, 95 (1993).

\bibitem{boyer}
        L. L. Boyer, J. Phys. C {\bf 17}, 1825 (1984).

\bibitem{murtaza}
        G. Murtaza, I. Ahmad and A. Afaq,
        Solid State Sciences {\bf 16}, 152 (2013).

\bibitem{ephraim}
        K. Ephraim Babu, A. Veeraiah, D. Tirupati Swamy and V. Veeraiah,
        Chin. Phys. Lett. {\bf 29}, 117102 (2012).
				
\bibitem{meziani2012}
        A. Meziani and H. Belkhir, Comput. Mater. Sci. {\bf 61}, 67 (2012).

\bibitem{brik}
        C.-G. Ma and M. G. Brik, Comput. Mater. Sci. {\bf 58}, 101 (2012).

\bibitem{sato2002}
        H. Sato, K. Shimamura, A. Bensalah, N. Solovieva, A. Beitterova, A. Vedda, M. Matine, H. Machida,
        T. Fukuda and M. Nikl, Jpn. J. Appl. Phys. {\bf 41} 2028 (2002).

\bibitem{benziada1984}
         A. Benziada-Taibi, J. Ravez and P. Hagenmuller,
         J. Fluorine Chem. {\bf 26}, 395 (1984).

\bibitem{zahn} D. Zahn, S. Herrmann and P. Heitjans,
         Phys. Chem. Chem. Phys. {\bf 13}, 21492 (2011).

\bibitem{boumriche1989}
         A. Boumriche, P. Simon, M. Rousseau, J. Y. Gesland and F. Gervais,
         J. Phys.: Condens. Matter {\bf 1}, 5613 (1989).
				
\bibitem{boumriche1994}
         A. Boumriche, J. Y. Gesland, A. Bulou, M. Rousseau, J. L. Fourquet and B. Hennion,
         Solid State Commun. {\bf 91}, 125 (1994).
				
\bibitem{mortier}
         M. Mortier, B. Piriou, J. Y. Buza\'re, M. Rousseau and J. Y. Gesland,
         Phys. Rev. B {\bf 67}, 115126 (2003).


\bibitem{mahlik}
         S. Mahlik, M. Grinberg, L. Shi and H. J. Seo,
         J. Phys.: Condens. Matter {\bf 21}, 235603 (2009).

 \bibitem{daniel}
 C. Cazorla and D. Errandonea, Phys. Rev. Lett. {\bf 113}, 235902 (2014).

\bibitem{mishra}
         A. K. Mishra, N. Garg, K. V. Shanavas, S. N. Achary, A. K. Tyagi and S. M. Sharma,
         J. Appl. Phys. {\bf 110} 123505 (2011).
				
\bibitem{vaitheekmgf3}
         G. Vaitheeswaran, V. Kanchana, R. S. Kumar, A. L. Cornelius, M. F. Nicol, A. Svane,
         A. Delin and B. Johansson,
				 Phys. Rev. B {\bf 76}, 014107 (2007).
				
\bibitem{vaitheecscdf3}
         G. Vaitheeswaran, V. Kanchana, R. S. Kumar, A. L. Cornelius, M. F. Nicol, A. Svane,
         N. E. Christensen and O. Eriksson,
				 Phys. Rev. B {\bf 81}, 075105 (2010).

\bibitem{oka}
         O. K. Andersen, Phys. Rev. B {\bf 12}, 3060 (1975).

\bibitem{savrasov}
         S. Y. Savrasov. Phys. Rev. B {\bf 54}, 16470 (1996).

\bibitem{vosko}
         S. H. Vosko, L. Wilk and M. Nusair, Can. J. Phys. {\bf 58},  1200 (1980).

\bibitem{perdew}
         J. P. Perdew, K. Burke, and M. Ernzerhof, Phys. Rev. Lett. {\bf 77}, 3865 (1996).

\bibitem{birch}
         F. Birch, Phys. Rev. {\bf 71}, 809 (1947).

\bibitem{oxides}
         V. Kanchana, G. Vaitheeswaran, A. Svane and A. Delin,
         J. Phys.: Condens. Matter, {\bf 18}, 9615 (2006).

\bibitem{giannozzi}
         P. Giannozzi, S. De Gironcoli, P. Pavone and S. Baroni,
         Phys. Rev. B {\bf 43},  7231 (1991).

\bibitem{baroni}
         P. Giannozzi et al Journal of Physics: Condensed Matter  21, 395502 (2009).

\bibitem{monkhorst}
        H. J. Monkhorst and J. D. Pack, Phys. Rev. B {\bf 13}, 5188 (1976).

\bibitem{Kotani1} M. van Schilfgaarde, T. Kotani and S. Faleev, Phys. Rev. Lett. \textbf{96}, 226402 (2006)

\bibitem{kotani}
        T. Kotani, M. van Schilfgaarde and S. V. Faleev, Phys. Rev. B {\bf 76}, 165106 (2007).

\bibitem{methfessel}
        M. Methfessel, M. van Schilfgaarde and R. A. Casali,
        in \emph{Lecture Notes in Physics}, edited by H. Dreysse (Springer-Verlag, Berlin, 2000), Vol. 535, p. 114.

\bibitem{Chantis} A. N. Chantis, M. van Schilfgaarde and T. Kotani, Phys. Rev. Lett. \textbf{96}, 086405 (2006).

\bibitem{Chantis-ss} A. N. Chantis, M. Cardona, N. E. Christensen, D. L. Smith, M. van Schilfgaarde,
          T. Kotani, A. Svane and R. C. Albers, Phys. Rev. B \textbf{78}, 075208 (2008).


\bibitem{gesland1972}
        J. Y. Gesland, M. Binois and J. Nouet, C. R. Acad. Sci. Paris {\bf 275B}, 551 (1972).

\bibitem{jiang2006}
        L. Q. Jiang, J. K. Guo, H. B. Liu, M. Zhou, X. Zhou, P. Wu and C. H. Li,
				J. Phys. Chem. Solids {\bf 67}, 1531 (2006).

\bibitem{moreira2007}
        R. L. Moreira, A. Dias, J. Phys. Chem. Solids {\bf 68}, 1617 (2007).

\bibitem{verma2009}
        A. S. Verma and V. K. Jindal, J. Alloys. Compds. {\bf 485}, 514 (2009).

\bibitem{luana1997}
        V. Luana, A. Costales, A. M. Pendas, M. Florez and V. M. G. Fernandez,
				Solid State Commun. {\bf 104}, 47 (1997).

\bibitem{brik2012}
        M. G. Brik, G. A. Kumar and D. K. Sardar, Mat. Chem. Phys. {\bf 136}, 90 (2012).

\bibitem{ridou1986}
        C. Ridou, M. Rousseau and F. Gervais, J. Phys. C {\bf 19}, 5757 (1986).
				
\bibitem{korba2009}
        S. A. Korba, H. Meradji, S. Ghemid and B. Bouhafs, Comput. Mater. Sci. {\bf 44}, 1265 (2009).
			























\end{thebibliography}
\end{document}